\documentclass[final,technote,10pt]{IEEEtran}

\usepackage{amsmath,amsfonts,amssymb}
\usepackage{graphicx, wrapfig}
\usepackage{color}
\usepackage{float}
\usepackage{tabularx,colortbl}
\usepackage[normalem]{ulem}

\usepackage[font=footnotesize]{caption}
\usepackage{subcaption}

\newcommand{\ba}{\vec {\bf a}}
\newcommand{\bk}{\vec {\bf k}}
\newcommand{\bkn}{\hat {\bf k}}
\newcommand{\br}{\vec {\bf r}}
\newcommand{\bu}{\vec {\bf u}}

\newcommand{\bE}{\vec {\bf E}}
\newcommand{\bH}{\vec {\bf H}}
\newcommand{\bP}{\vec {\bf P}}
\newcommand{\bS}{\vec {\bf S}}

\newcommand{\bq}{\vec {\bf q}}
\newcommand{\bF}{\vec {\bf F}}

\newcommand{\bA}{\vec {\bf A}}
\renewcommand{\arraystretch}{1.5} 

\newcommand{\jmpE}{\left[\hspace{-0.25em}\left[ \vec {\bf E}\right]\hspace{-0.25em}\right]}
\newcommand{\jmpH}{\left[\hspace{-0.25em}\left[ \vec {\bf H}\right]\hspace{-0.25em}\right]}
\newcommand{\jmpP}{\left[\hspace{-0.25em}\left[ \vec {\bf P}\right]\hspace{-0.25em}\right]}
\newcommand{\jmpS}{\left[\hspace{-0.25em}\left[ \vec {\bf S}\right]\hspace{-0.25em}\right]}

\begin{document}

\title{A Discontinuous Galerkin Time Domain Framework for Periodic Structures Subject To Oblique Excitation}

\author{Nicholas C. Miller, Andrew D. Baczewski, John D. Albrecht, and Balasubramaniam Shanker}

\maketitle

\begin{abstract}

A nodal Discontinuous Galerkin (DG) method is derived for the analysis of time-domain (TD) scattering from doubly periodic PEC/dielectric structures under oblique interrogation.  Field transformations are employed to elaborate a formalism that is free from any issues with causality that are common when applying spatial periodic boundary conditions simultaneously with incident fields at arbitrary angles of incidence.  An upwind numerical flux is derived for the transformed variables, which retains the same form as it does in the original Maxwell problem for domains without explicitly imposed periodicity.   This, in conjunction with the amenability of the DG framework to non-conformal meshes, provides a natural means of accurately solving the first order TD Maxwell equations for a number of periodic systems of engineering interest.  Results are presented that substantiate the accuracy and utility of our method.

\end{abstract}


\section{Introduction}

\begin{figure}[!b]
	\centering
	\includegraphics[width=0.325\textwidth]{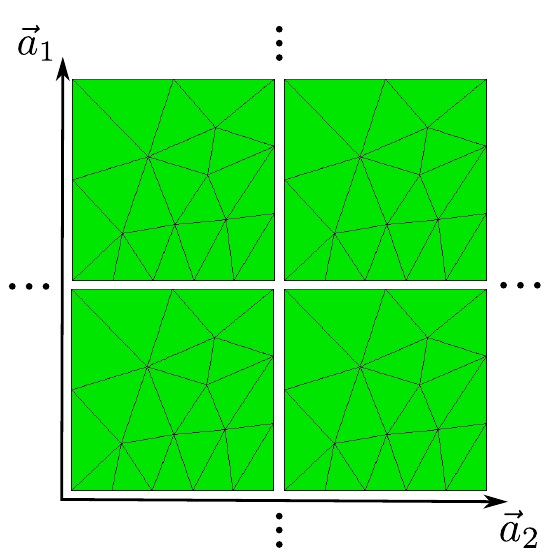}
	\caption{Illustration of the $z$-plane of a doubly periodic structure with periods $|\ba_1|$ and $|\ba_2|$.  The ellipses indicate that the structure is periodic in the $x$- and $y$-directions.}
	\label{fig:unitcells}
\end{figure}

\IEEEPARstart{P}{eriodic} structures play a significant role in electromagnetics and optics in generating unique spectral responses that can be readily engineered.  Applications of periodicity include frequency selective surfaces (FSS) \cite{Munk2005}, electromagnetic band gap (EBG) structures \cite{Yang2007}, biomimetic structures and metamaterials \cite{Munk2009}, \cite{Capolino2010}, etc.  Computational analysis of fields in increasingly intricate periodic unit cells plays a significant role in their design and optimization.  In the frequency domain, Integral Equation (IE) \cite{Baczewski2012}, \cite{Baczewski2012a}, Finite Element (FE) \cite{Lucas1995}, \cite{Sotirelis2007}, and Discontinuous Galerkin (DG) \cite{Chun2008} methods have been successfully applied to a variety of periodic electromagnetic systems. Time-domain (TD) methods for studying periodic systems include FE \cite{Petersson2006},\cite{Petersson2006a}, IE \cite{Chen2005}, and Finite Difference Time Domain (FDTD) \cite{Harms1994}, while DG methods remain relatively unexplored.

TD analysis of periodic structures provides a number of advantages, such as characterization of the broadband response of a structure in a single simulation, and treatment of nonlinearities.  Both integral and differential formulations of the Maxwell problem have attendant disadvantages as well.  For integral formulations, discretization yields a dense linear system.  While fast and efficient \cite{Chen2005}, \cite{Dault2012} methods have been applied to these problems, stable formulations of TDIEs remain a research problem, with much recent progress \cite{Pray2012}.  Recent work has also been presented on obtaining transient response using entire domain Laguerre polynomials that results a system wherein the time variable is completely avoided \cite{Jung2011}.  Alternatively, while differential formulations of the problem yield sparse linear systems and stability is better understood, the proper imposition of boundary conditions (BCs) becomes challenging.  In particular, the asymptotic boundary condition on the fields receding to infinity must be enforced approximately with an absorbing boundary condition (ABC) or a perfectly matched layer (PML) \cite{Jin2010a}.  Further, while periodic BCs at the perimeter of the unit cell are trivial to enforce for systems excited at normal incidence, there are well-known issues associated with causality at oblique incidence \cite{Petersson2006}.

A set of field transformations that mitigate causality issues was introduced for FDTD in 1993 \cite{Veysoglu1993}, and later adapted to an FETD framework in a sequence of papers in the mid-2000s \cite{Petersson2006}, \cite{Petersson2006a}. Here, the frequency domain Floquet-periodic boundary condition is exploited, wherein fields at the unit cell boundaries are related to one another by a phase shift that depends on the exciting wave vector and lattice vectors.  The frequency domain Maxwell Equations are then posed in terms of a set of transformed variables, into which this phase shift is built, and an inverse transform is applied to return the equations to the time domain. Additional terms then appear in the TD Maxwell Equations for the transformed variables.

In this work, we will apply these field transformations to a time domain Discontinuous Galerkin (DG) framework for the conservation form of the Maxwell equations for the first time.  Time domain analysis of periodic structures with DG methods has received relatively little attention, with a few exceptions \cite{Chun2008}, \cite{Sirenko2013}.  The unique contributions of this paper are extensions of a time domain DG framework that permit the analysis of doubly periodic structures at oblique incidence.  First, the field transformations that are used to remove causality issues are reviewed.  We then demonstrate that the form of the upwind flux utilized in discretizing the transformed Maxwell Equations is invariant to whether or not one is utilizing the original or transformed fields. Issues addressing the use of non-conformal meshes across periodic boundaries are discussed, and relevant implementation details are provided.  Finally, results are presented that validate the accuracy and utility of our method for a number of doubly periodic test cases.

\section{Mathematical Formulation}

\begin{figure}[!h]
	\centering
	\includegraphics[width=0.3\textwidth]{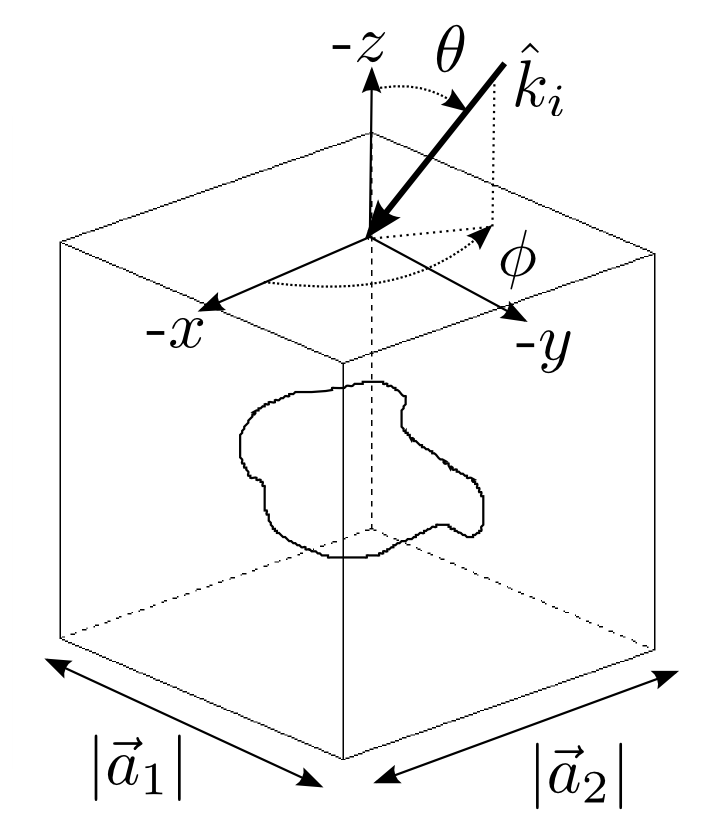}
	\caption{Illustration of a single unit cell of a doubly periodic structure with periods $|\ba_1|$ and $|\ba_2|$.}
	\label{fig:single_unitcell}
\end{figure}

Consider a domain, $\Omega \subset \mathbb{R}^3$ depicted in Fig. \ref{fig:single_unitcell}, where a doubly periodic distribution of isotropic, lossless, dielectric and/or PEC scatterers reside.  The periodicity of the system is described by a 2-lattice, $\mathcal{L}_2$, defined as: 
\begin{equation}
 \mathcal{L}_2 = \lbrace \bu_n = n_1 \ba_1 + n_2 \ba_2 | n_1, n_2 \in \mathbb{Z}\rbrace 
\end{equation}
Here, the subscript $n$ is defined as a multi-index, and $\ba_i$ are basis vectors for the lattice.  These vectors will be orthogonal in this work, but extensions to non-orthogonal basis vectors are simply realized.  Incident on the system is a planewave excitation $\bE_{i}(\br,t)$, with a wavevector $\bkn_{i}=\sin\theta\cos\phi\hat{x}+\sin\theta\sin\phi\hat{y}+\cos\theta\hat{z}$.  The incident wavevector, $\bkn_{i}$, can be further decomposed into $\bkn_{i}^\parallel$ and $\bkn_{i}^{\perp}$, which are within and orthogonal to the span of $\mathcal{L}_2$, respectively.

The fields obey the following boundary conditions under spatial translation by a lattice vector in $\mathcal{L}_2$:
\begin{subequations}
\begin{align}
 \bE(\br,t) = \bE(\br+\bu_n,t) \ast \delta\left(t+\frac{\bkn_i^{\parallel}\cdot\br}{c_0}\right) \label{eq:td_notrans_perbc_a}\\
 \bH(\br,t) = \bH(\br+\bu_n,t) \ast \delta\left(t+\frac{\bkn_i^{\parallel}\cdot\br}{c_0}\right) \label{eq:td_notrans_perbc_b}
\end{align}
\end{subequations}
Direct implementation of these periodic boundary conditions requires knowledge of future values of fields at one periodic boundary in order to update fields at the other periodic boundary.  In the context of a time integration scheme in which fields are updated in time based upon a sequence of their previous values, this is not possible without extrapolation.  

Alternatively, transformed fields can be identified for which the periodic boundary conditions remain causal.  As done in \cite{Petersson2006},\cite{Veysoglu1993},  we introduce delayed auxiliary variables, $\bP(\br,\omega)$ and $\bS(\br,\omega)$ 

\begin{subequations}
\begin{align}
 \bE(\br,\omega) = \bP(\br,\omega) e^{-j\bk_i^{\parallel}\cdot\br} \label{eq:fd_trans_a}\\
 \bH(\br,\omega) = \bS(\br,\omega) e^{-j\bk_i^{\parallel}\cdot\br} \label{eq:fd_trans_b}
\end{align}
\end{subequations}
It can be shown trivially that these transformed fields obey 
\begin{subequations}
\begin{align}
 \bP(\br,t) = \bP(\br+\bu_n,t) \label{eq:periodic_bcs_P}\\
 \bS(\br,t) = \bS(\br+\bu_n,t) \label{eq:periodic_bcs_S}
\end{align}
\end{subequations}
As is evident from Eqns. (\ref{eq:periodic_bcs_P}) and (\ref{eq:periodic_bcs_S}), using these auxiliary field components is tantamount to zero phase propagation at the boundaries, i.e., there is no delay in boundaries of the unit cell.  This is the time domain analog to cell-periodic Bloch functions typical of frequency analysis.

Applying the field transformations to the first order time domain Maxwell Equations yields
\begin{subequations}
\begin{align}
\varepsilon\frac{\partial \bP(\br,t)}{\partial t} + \frac{\bkn_{i}^\parallel}{c_0} \times \frac{\partial\bS(\br,t)}{\partial t}  &=  \nabla \times \bS(\br,t) \label{pme_curlS}\\
- \frac{\bkn_{i}^\parallel}{c_0} \times \frac{\partial\bP(\br,t)}{\partial t} + \mu\frac{\partial \bS(\br,t)}{\partial t}   &=-\nabla \times \bP(\br,t) \label{pme_curlP}
\end{align}
\end{subequations}
It is these equations that we will now discretize within the DG framework.

\section{The Discontinuous Galerkin Method}

\subsection{Discretization}

To allow a seamless extension from previous DG formulations \cite{Hesthaven2002}, \cite{Niegemann2009}, \cite{Busch2011}, we write Eqns. (\ref{pme_curlS}) and (\ref{pme_curlP}) in conservation form:
\begin{equation}
Q \frac{\partial \bq(\br,t)}{\partial t} + \nabla \cdot \bF\left(\bq(\br,t)\right) = 0 \label{eq:cons_eq}
\end{equation}
Here, the periodic/materials matrix $Q$, field six-vector $\bq(\br,t)$, and flux matrix $\bF\left(\bq(\br,t)\right)$ are defined as:

\begin{equation*}
Q = \left( \begin{array}{cc}
\varepsilon \mathcal{I}_1  & c_0^{-1} \bkn_{i}^\parallel \times \mathcal{I}_1  \\
- c_0^{-1} \bkn_{i}^\parallel \times \mathcal{I}_1 & \mu \mathcal{I}_1
\end{array} \right),
\end{equation*}

\begin{equation*}
\bq(\br,t) = \left( \begin{array}{c} \bP(\br,t) \\ \bS(\br,t) \end{array} \right),
\bF\left(\bq(\br,t)\right) = \left( \begin{array}{c} - \hat{e}_i \times \bS(\br,t) \\ \hat{e}_i \times \bP(\br,t) \end{array} \right)
\end{equation*}
here, $\hat{e}_i$ represents the i$^{th}$ Cartesian unit vector, $\varepsilon$ is the isotropic permittivity, $\mu$ is the isotropic permeability, and $\mathcal{I}_1$ is the 3x3 identity matrix.  

Solving this system of equations requires discretizing the domain using $k$ non-overlapping tetrahedra, where domains are denoted $\Omega^k$ with boundaries $\partial\Omega^k$ that are equipped with an outward pointing normal $\hat{n}$.  The vector unknowns are expanded into a set of globally discontinuous nodal polynomials $\bq\left(\br,t \right) \approx \sum\limits_{i=1}^{N_p} \bq^k\left(\br_i,t\right) \ell_i^k \left(\br\right) $.  We use the nodal basis functions defined in \cite{Hesthaven2002}.

Following standard DG practice \cite{Hesthaven2002}, a strong form of the problem is obtained as:
\begin{align}
\iiint\limits_{\Omega^k} \left(Q \frac{\partial\bq(\br, t)}{\partial t} + \nabla \cdot \bF\left(\bq(\br, t)\right)\right)\ell_j^k (\br)d\br \nonumber \\
= \iint\limits_{\partial\Omega^k} \vec{n} \cdot \left( \bF\left(\bq(\br, t)\right) - \bF^*\left(\bq(\br, t)\right) \right)\ell_j^k (\br) d\br \label{eq:strong_form}
\end{align}
where $\bF^*$ is called the numerical flux.  We can rewrite the semi-discrete problem in Eqn. (\ref{eq:strong_form}) as:

\begin{equation}
\frac{\partial \bq(\br,t)}{\partial t} = Q^{-1} \left(\mathcal{M}^{-1} \mathcal{S}\bq + \mathcal{M}^{-1}\mathcal{F} \left[ \hat{n} \cdot \left(\bF - \bF^* \right)\right]\right) \label{eq:semidiscrete}
\end{equation}
with the function of nodal values $\hat{n}\cdot\left(\bF - \bF^* \right)$, defined on the element boundaries, replacing the flux matrix $\bF\left(\bq(\br,t)\right)$, the periodic/materials matrix $\mathcal{Q}$ re-defined as

\begin{equation*}
Q = \left( \begin{array}{cccccc}
\varepsilon \mathcal{I}_2  & 0 & 0 & 0 & 0 & -\kappa_y  \mathcal{I}_2 \\
0 & \varepsilon \mathcal{I}_2 & 0 & 0 & 0 & \kappa_x  \mathcal{I}_2 \\
0 & 0 & \varepsilon \mathcal{I}_2 & \kappa_y \mathcal{I}_2 & -\kappa_x  \mathcal{I}_2 & 0 \\
0 & 0 & \kappa_y \mathcal{I}_2 & \mu \mathcal{I}_2  & 0 & 0\\
0 & 0 &  -\kappa_x \mathcal{I}_2 & 0 & \mu \mathcal{I}_2 & 0\\
-\kappa_y \mathcal{I}_2 & \kappa_x \mathcal{I}_2 & 0 & 0 & 0 & \mu \mathcal{I}_2
\end{array} \right)
\end{equation*}
where $\bkn_{i}^\parallel = \kappa_x\hat{x} + \kappa_y \hat{y}$ and $\mathcal{I}_2$ is the $N_p$x$N_p$ identity matrix.  The mass matrix $\mathcal{M}$, stiffness matrix $\mathcal{S}$, and face matrix $\mathcal{F}$ are defined as 

\begin{subequations}
\begin{align*}
\mathcal{M}_{ij} = \iiint\limits_{\Omega^k} \ell_i^k (\br) \ell_j^k (\br) d\br \\
\mathcal{S}_{ij} = \iiint\limits_{\Omega^k} \ell_i^k (\br) \nabla \ell_j^k (\br) d\br \\
\mathcal{F}_{ij} = \iint\limits_{\partial\Omega^k} \ell_i^k (\br) \ell_j^k (\br) d\br
\end{align*}
\end{subequations}

\subsection{Periodic Numerical Flux}

Choice of the nodal values $\hat{n} \cdot \left(\bF - \bF^* \right)$ is at the heart of all DG formulations.  Hesthaven and Warburton have proven that an upwind flux is both stable and convergent for Maxwell's Equations \cite{Hesthaven2002}.  For the non-periodic Maxwell's Equations, the upwind flux takes the form

\begin{equation}
\hat{n} \cdot \left( \bF - \bF^* \right) = \left( \begin{array}{c} -\bar{Z}^{-1} \hat{n} \times \left(Z^+ \jmpH - \hat{n} \times \jmpE \right)\vspace{0.25cm} \\  \bar{Y}^{-1} \hat{n} \times \left(Y^+ \jmpE + \hat{n} \times \jmpH \right) \end{array} \right) \label{eq:nonper_flux}
\end{equation}
Here, the jump $\jmpE=\bE^+ - \bE^-$ is defined in terms of nodal field values at the element boundaries, and the impedance $\bar{Z}=Z^+ + Z^-$ is twice the average impedance shared at these boundaries.
To derive the periodic numerical flux for $\bP(\br,t)$ and $\bS(\br,t)$, we note that $\bE=\bP\ast \delta\left(t-\frac{\bkn_i^{\parallel}\cdot \vec{r}}{c_0}\right)$ and $\bH=\bS\ast \delta\left(t-\frac{\bkn_i^{\parallel}\cdot \vec{r}}{c_0}\right)$.  Using these in the conservation form of Maxwell's equations

\begin{eqnarray*}
\left( \begin{array}{cc}
\varepsilon \mathcal{I}_1 & 0 \\
0 & \mu \mathcal{I}_1  
\end{array} \right) \frac{\partial}{\partial t}
\left( \begin{array}{c} \bP\ast \delta\left(t-\frac{\bkn_i^{\parallel}\cdot \vec{r}}{c_0}\right) \\ \bS\ast \delta\left(t-\frac{\bkn_i^{\parallel} \cdot \vec{r}}{c_0}\right) \end{array} \right) \\
+ \nabla \cdot \left( \begin{array}{c} - \hat{e}_i \times \bS \ast \delta\left(t-\frac{\bkn_i^{\parallel} \cdot \br}{c_0}\right) \\ \hat{e}_i \times \bP \ast \delta\left(t-\frac{\bkn_i^{\parallel} \cdot \br}{c_0}\right) \end{array} \right) = 0
\end{eqnarray*}
it is evident that this system has two distinct characteristic values, $\pm \left(\varepsilon\mu \right)^{-1/2}$.  This implies that only three Rankine-Hugoniot jump conditions are needed to relate the fields across discontinuities \cite{Hesthaven2002}, \cite{LeVeque2002}.  Using the convention in \cite{Mohammadian1991}, integrating over a single element, and reducing integration limits to the faces of the elements, we arrive at the jump conditions for the equivalent transformed equations

\begin{equation*}
\left[ Z^-\left(\bS^* - \bS^-\right) + \hat{n}\times\left(\bP^* - \bP^-\right)\right]\ast \delta\left(t-\frac{\bkn_i^{\parallel}\cdot\br}{c_0}\right) = 0
\end{equation*}

\begin{equation*}
\left[ Z^+\left(\bS^{**} - \bS^+\right) + \hat{n}\times\left(\bP^{**} - \bP^+\right)\right]\ast \delta\left(t-\frac{\bkn_i^{\parallel}\cdot\br}{c_0}\right) = 0
\end{equation*}

\begin{equation*}
\left[\hat{n}\times\left(\bP^{**} - \bP^{*}\right)\right]\ast \delta\left(t-\frac{\bkn_i^{\parallel}\cdot\br}{c_0}\right) = 0
\end{equation*}

\begin{equation*}
\left[\hat{n}\times\left(\bS^{**} - \bS^{*}\right)\right]\ast \delta\left(t-\frac{\bkn_i^{\parallel}\cdot\br}{c_0}\right) = 0
\end{equation*}
Since these equations hold for all time, the periodic numerical flux may now be written as \cite{Mohammadian1991}

\begin{equation}
\hat{n} \cdot \left( \bF - \bF^* \right) = \left( \begin{array}{c} -\bar{Z}^{-1} \hat{n} \times \left(Z^+ \jmpS - \hat{n} \times \jmpP \right) \vspace{0.25cm} \\ \bar{Y}^{-1} \hat{n} \times \left(Y^+ \jmpP + \hat{n} \times \jmpS \right) \end{array} \right) \label{eq:per_flux}
\end{equation}
In Eqn. \ref{eq:per_flux}, $\jmpP = \bP^+ - \bP^-$ is the jump in the nodal field values at an element's boundaries.

\subsection{Boundary Conditions}

\begin{table}[!h]

\renewcommand{\arraystretch}{2.0}
\caption{Boundary Condition Jumps}
\label{jmp_table}
\centering

\begin{tabular}{|c||c||c|}
\hline
B.C.  &  $\jmpP$  & $\jmpS$ \\ 
\hline
PEC:		&	$-2\bP^-$		&	0\\ 	
\hline
ABC (TE):		&	$-2\bP^-\left|\cos\theta\right|$	&	$-2\bS^-$ \\
\hline
ABC (TM):		&	$-2\bP^-$	&	$-2\bS^-\left|\cos\theta\right|$ \\
\hline
TF/SF:	&	$\bP^+-\bP^- \pm \bP^{inc}$	&  $\bS^+-\bS^- \pm \bS^{inc}$\\     	
\hline
\end{tabular}
\end{table}

Applying boundary conditions to the periodic system of equations requires constraining the jumps \vspace{0.1cm} $\jmpP$ and $\jmpS$ across a face.  We present a list of common DG jumps first presented in \cite{Niegemann2009}.  Here, TF/SF denotes total fields and scattered fields, respectively.  The addition of the angle of incidence in the jumps for the planewave ABC allows the periodic numerical flux to satisfy the well-known Silver-M\"{u}ller condition for the transformed fields
\begin{eqnarray*}
	Z\hat{n} \times \bS = \left|\cos\theta\right| \hat{n} \times \hat{n} \times \bP \\
	Y\hat{n} \times \bP =  -\left|\cos\theta\right| \hat{n} \times \hat{n} \times \bS \label{eq:sm_cond}
\end{eqnarray*}
for TE and TM polarization, respectively.  Here, $Z=1/Y$ is the impedance of the medium.

We must also consider boundary conditions on the interfaces between unit cells.  To implement Eqns. (\ref{eq:periodic_bcs_P}) and (\ref{eq:periodic_bcs_S}), a map must be created between the periodic planes of the unit cell.  A natural first choice for creating these maps is to create a meshed unit cell in which the periodic planes are conformal, and set the jumps to be $\jmpP = \bP(\br+\bu_n,t) - \bP(\br,t)$ and $\jmpS = \bS(\br+\bu_n,t) - \bS(\br,t)$.  Alternatively, it is significantly easier to generate a meshed unit cell without meticulous constraints on the periodic planes.  The nodes of the periodic plane will not align, and information regarding the non-conformal triangles is generated.
This interface is first decomposed into a list of the four different types of fragments:  three-, four-, five-, and six-vertex fragments.  A polygon clipping algorithm \cite{Vatti1992} is employed to generate this data.  These fragments are defined to facilitate the definition of quadrature rules for numerically integrating surface terms.
\begin{figure}[!b]
	\centering
	\includegraphics[width=0.49\textwidth]{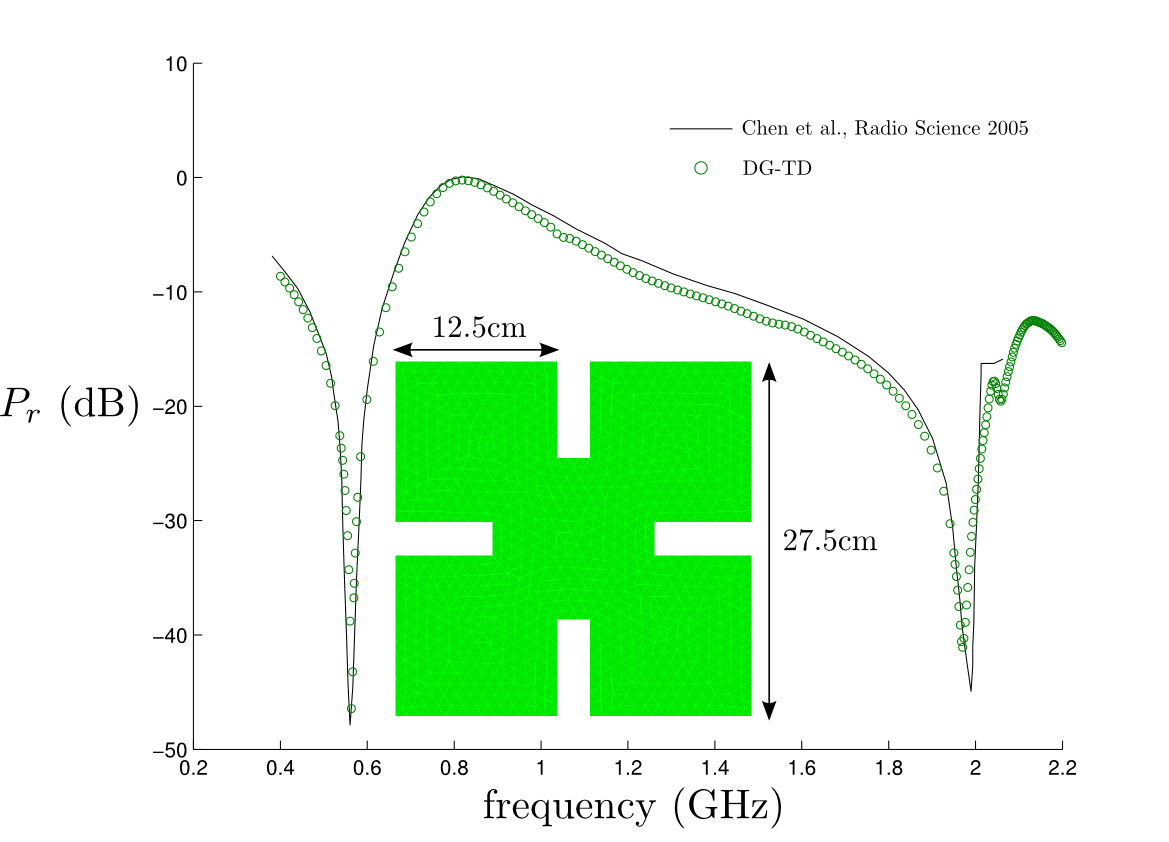}
	\caption{Reflection coefficient (in dB) of a planewave normally incident on periodically arranged PEC Minkowski Fractals.  The unit cell dimensions for the fractal are $|\ba_1|=|\ba_2|=30$cm.  Dimensions of the fractal are shown above.  The ABC surfaces were placed $10$cm away from the PEC fractal in $\pm z$-direction.    The electric field is $x$-polarized.}
	\label{fig:minfrac}
\end{figure}
\begin{figure}[!t]
	\includegraphics[width=0.49\textwidth]{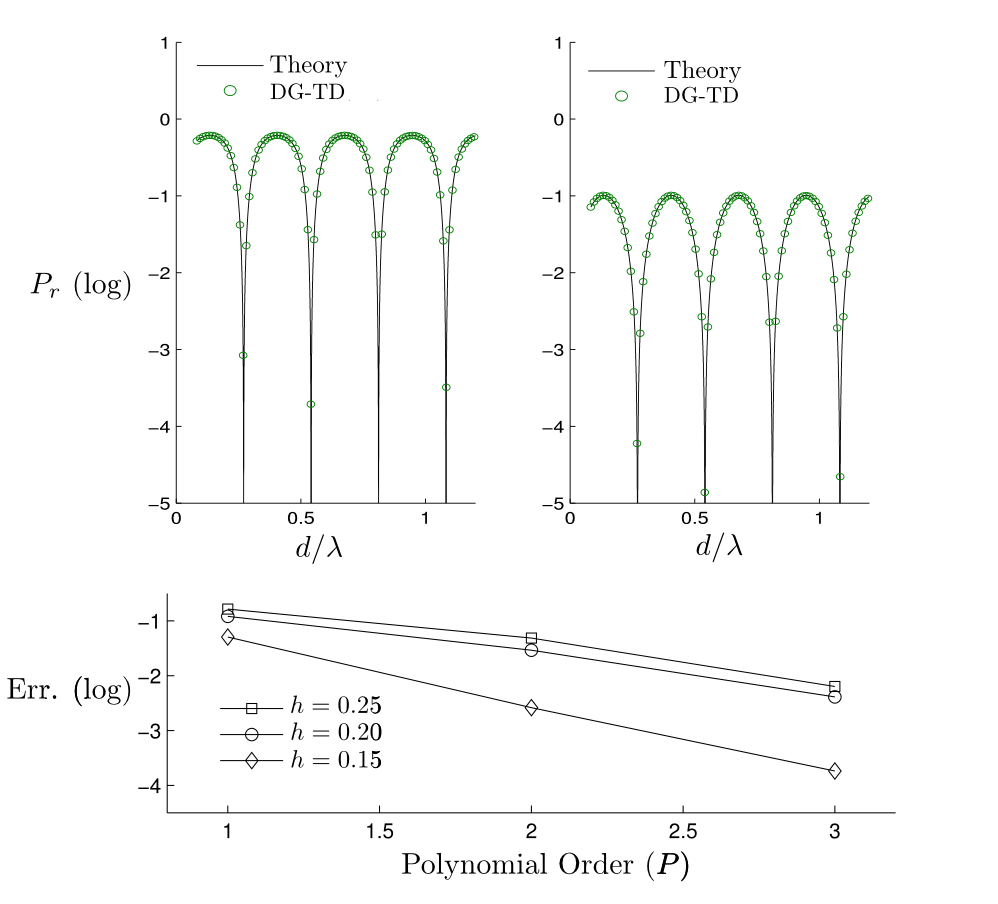}
	\caption{Power reflected from a planewave obliquely incident on a nonmagnetic and lossless dielectric slab, $\theta=50^\circ$.  Top: Power reflection over broadband frequency range for TE polarization (top left) and TM polarization (top right).  Bottom:  minimum edge length ($h$) and polynomial order ($P$) error convergence for TE polarization.}
	\label{fig:slab}
\end{figure}

\section{Results}

To demonstrate the validity of our computational framework, we discuss several scattering results.    In all cases, a low-storage fourth order Runge-Kutta integration \cite{Carpenter1994} is used with a time step size determined by $c\Delta t=h P^{-2}$, where $h$ is the minimum edge length and $P$ is the polynomial order.  Reflection or transmission data presented for each structure is obtained from Eqn. (\ref{eq:power_ref}).

\begin{equation}
P_{r/t}(f) = \frac{\left| \bE_{r/t} (f) \right|^2}{\left| \bE_i (f) \right|^2} \label{eq:power_ref}
\end{equation}
Here, $\bE_i (f)$ is the Fourier transform of the planewave excitation.  The reflected and transmitted field, denoted by $\bE_{r/t} (f)$, is calculated as the magnitude of the Fourier transform of the fundamental coefficient $\bA_{00}(t)$ given as
\begin{equation}
\bA_{00}(t) = \frac{1}{|\ba_1||\ba_2|}\int\limits_{y=0}^{|\ba_2|}\int\limits_{x=0}^{|\ba_1|} \bP(x,y,z=z_{RT}; t)dxdy \label{eq:fund_coeff}
\end{equation}
This coefficient is integrated over the $z=z_{RT}$ plane \cite{Petersson2006} located either below or above the scattering structure for reflection or transmission, respectively.
\begin{figure}[!b]
	\centering
	\includegraphics[width=0.49\textwidth]{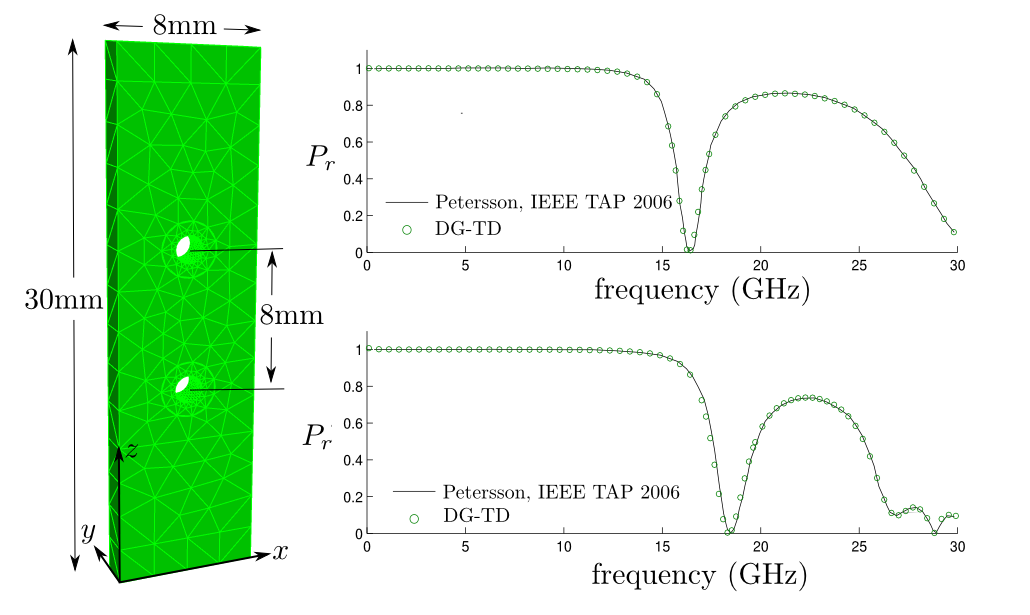}
	\caption{(left) Illustration of the PEC rods oriented in the $y$-direction.  The unit cell dimensions are $|\ba_1|=8$mm and $|\ba_2|=2$mm.  The radius of both rods is $0.8$mm. (right) Power reflected from a normally (top) and obliquely (bottom, $\theta=30^\circ$) incident planewave.  The electric field is $y$-polarized for both cases.}
	\label{fig:pec_rods}
\end{figure}

The first result is scattering of a plane wave normally incident on a Minkowski fractal FSS.  This result validates our implementation at normal incidence, and serves as a check of the non-conformal treatment of periodic boundary conditions independent of the oblique incidence framework.  Fig. \ref{fig:minfrac} displays an illustration of the fractal and its dimensions, and the unit cell dimensions were $|\ba_1|=|\ba_2|=30$cm.  An air box was placed above and below the PEC fractal with heights of $10$cm.  The DG-TD numerical results are displayed in Fig. \ref{fig:minfrac}.  Reference data for the Minkowski fractal was drawn from \cite{Chen2005}. 

The next structure is a simple dielectric slab of thickness $d=1.0$m and relative permittivity $\varepsilon_r=4.0$.  This slab is lossless and nonmagnetic.  The unit cell dimensions were chosen arbitrarily to be $|\ba_1|=|\ba_2|=0.35$m.  The height of the air box above and below the slab was chosen to be $1.0$m.  Fig. \ref{fig:slab} displays the power reflected from the slab with the angle of incidence $\theta=50^\circ$.  For this structure, we show excellent agreement between the theoretical and numerical power reflection coefficient across the frequency range.  To demonstrate 
\begin{figure}[!t]
	\centering
	\includegraphics[width=0.49\textwidth]{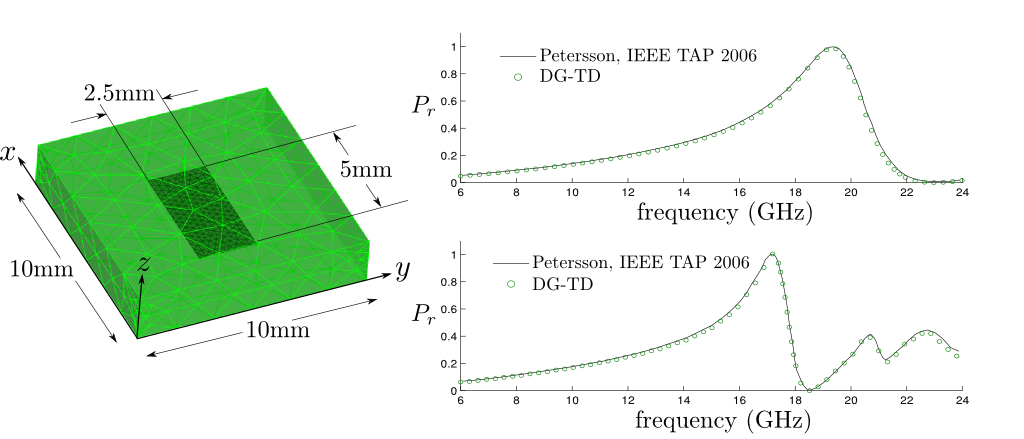}
	\caption{(left) Illustration of the nonmagnetic and lossless dielectric slab with periodically arranged PEC strips located at the center of the slab.  The slab has a thickness of $2$mm, and the PEC strips are $2.5$mm by $5$mm, as shown in the illustration. (right) Power reflected from a normally (top) and obliquely (bottom, $\theta=30^\circ$) incident planewave on a nonmagnetic and lossless dielectric slab with periodically arranged PEC strips residing at the center of the slab's thickness.  The electric field is $y$-polarized for both cases.}
	\label{fig:slab_pec}
\end{figure}
\begin{figure}[!b]
	\centering
	\includegraphics[width=0.49\textwidth]{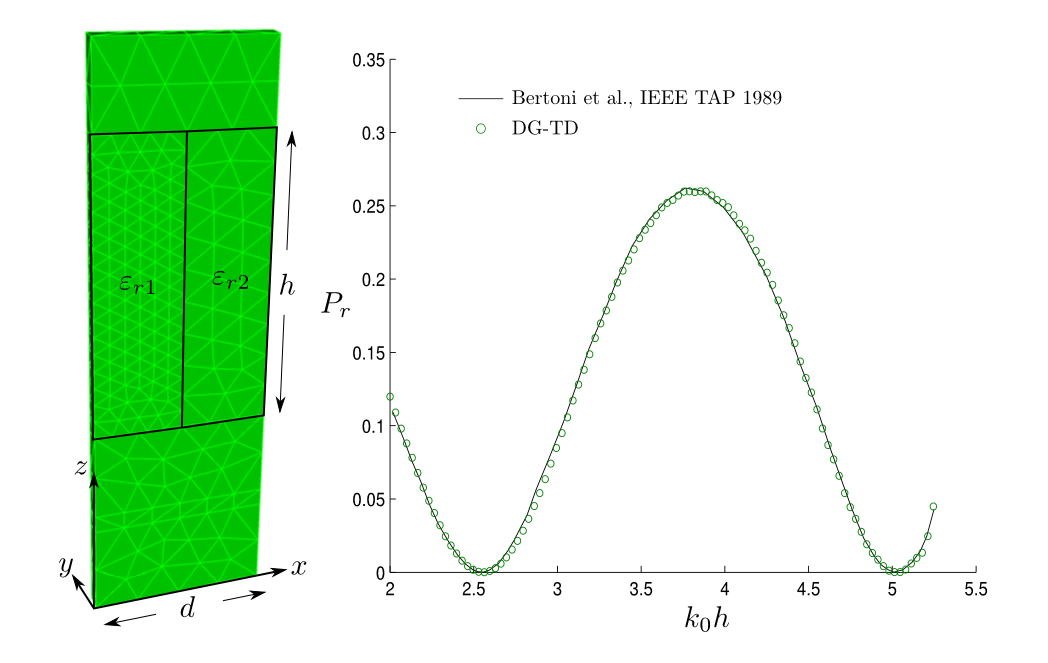}
	\caption{(left) Illustration of a single unit cell of periodically arranged dielectric slabs (outlined in black) in the $x$-direction with $\varepsilon_{r1}=2.56$ and $\varepsilon_{r2}=1.44$.  The slab heights and widths were chosen based on the ratio $h/d=1.713$ and $d/2.0$, respectively. (right) Reflected power of an obliquely incident planewave ($\theta=45^\circ$).  The electric field is $y$-polarized.}
	\label{fig:stacked_slabs}
\end{figure}
the higher order accuracy of the computational framework, Fig. \ref{fig:slab} displays the average absolute error between the numerically and theoretically calculated reflection over the frequency band.  

The next structure consists of two infinite PEC rods oriented in the $y$-direction.  The unit cell dimensions, displayed in Fig. \ref{fig:pec_rods}, are $8$mm by $2$mm in the $x$- and $y$-direction, respectively.  Length of the structure in the $y$-direction was chosen to reduce the number of unknowns, as it is infinite in the $y$-direction.  The air boxes above and below the rods are $11$mm from the centers of the rods, and the centers of the rods were placed $8$mm apart.  The radius of both rods is $0.8$mm.  Fig. \ref{fig:pec_rods} displays the numerical results of the periodic DG-TD method compared against the numerical results of the periodic FEM-TD method.  Our framework demonstrates excellent results compared to the FEM-TD framework.  The effect of the planewave ABC past the next higher order Floquet mode is also captured.

Our next structure is an array of PEC strips embedded in a dielectric slab.  The dielectric slab is lossless and nonmagnetic, and the dimensions are shown in Fig. \ref{fig:slab_pec}.  An air box was placed above and below the dielectric slab with a height of $30$mm in the $\pm z$-direction.  Reference data \cite{Petersson2006} agrees very well with the numerical results of the DG-TD code shown in Fig. \ref{fig:slab_pec}.  Again we see the effect of the planewave ABC much like the FEM-TD framework \cite{Petersson2006}.

Our last validation structure consists of dielectric slabs with alternating dielectric constants.  The dielectric slabs are lossless and nonmagnetic, and the unit cell is displayed in Fig. \ref{fig:stacked_slabs}.  Slab heights $h$ and width of the slabs $d$ are set based on the ratio $h/d=1.713$, and each slab's width was set to $0.5d$.  An air box was placed above and below the set of slabs with an arbitrarily chosen height of $0.5d$ above and $d$ below.  The relative permittivity of each slab was $\varepsilon_{r1} = 2.56$ and $\varepsilon_{r2} = 1.44$.  Results for this structure are shown in Fig. \ref{fig:stacked_slabs}, with reference data drawn from \cite{Bertoni1989}.  Our results show good agreement with the reference data.  

\begin{figure}[!b]
	\centering
	\includegraphics[width=0.49\textwidth]{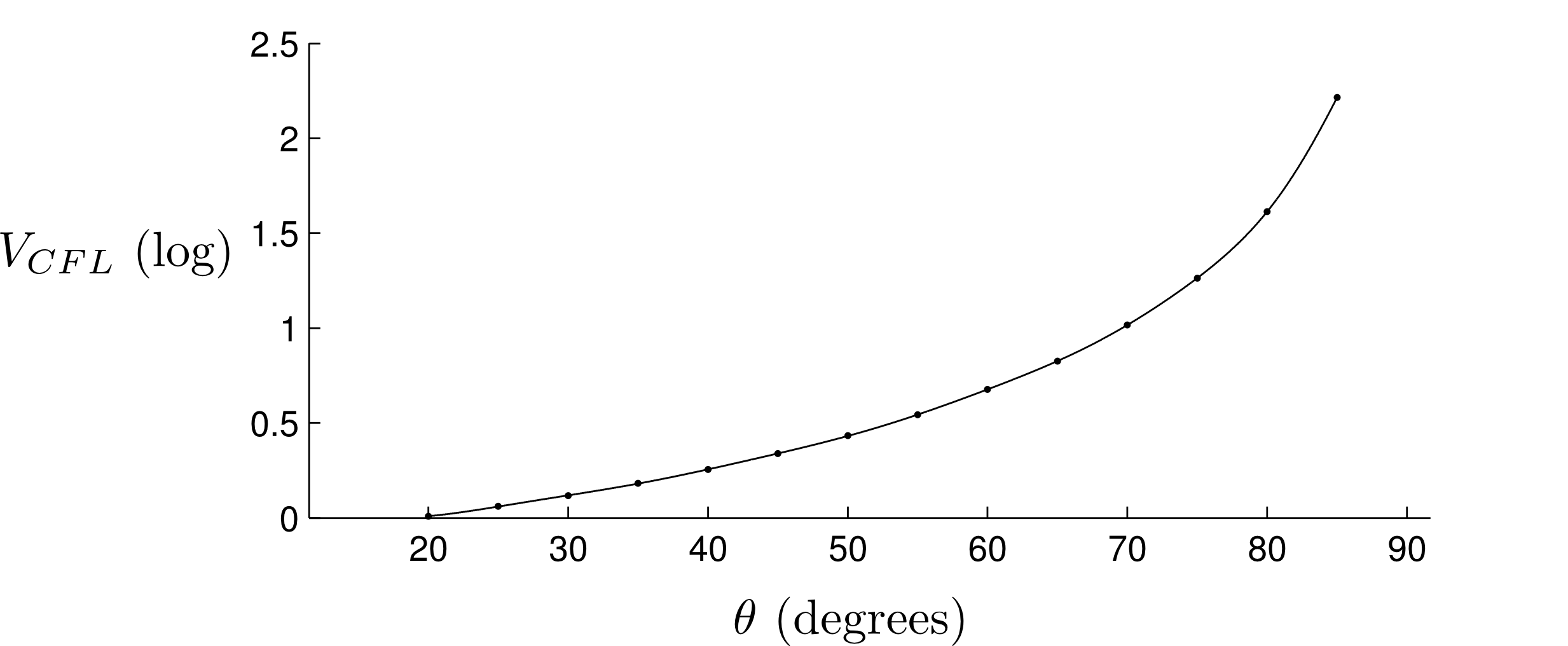}
	\caption{Angular dependence of time step scale $V_{CFL}$.  Angles less than $\theta=20^\circ$ required unity scaling for stability.}
	\label{fig:stability}
\end{figure}

We have shown several cases which validate this DGTD framework.  The final topic of this work is addressing the stability of the explicit time integrator with respect to the planewave's angle of incidence.  The speed of Floquet modes is proportional to $cos^{-1}\theta$ \cite{Petersson2006}, and therefore the CFL bound $c\Delta t \leq h P^{-2}$ is not sufficient for higher angles of incidence.  The simplest solution of this problem is to scale the CFL condition as $c\Delta t = h P^{-2} V_{CFL}^{-1}$.  Fig. \ref{fig:stability} displays the smallest stable time step scale with respect to angle of incidence for a planewave passing through freespace.  The unit cell dimensions for the freespace mesh were $|\ba_1| = |\ba_2| = \lambda_{min} / 2$, the smallest edge length was $h = \lambda_{min} / 10$, and the polynomial order was $P=2$.  These parameters were held constant for each angle of incidence.  The unit cell mesh was conformal with respect to the periodic boundaries.
This simple result provides empirical evidence that the explicit time integration scheme is conditionally stable, even at near grazing angles of incidence.  Satisfying the CFL condition at near grazing angles, however, requires scales of two orders of magnitude and thus increases the number of time steps accordingly.

\section{Conclusion and Future Work}

In this paper, we have presented a higher-order three-dimensional Time Domain Discontinuous Galerkin Method for analyzing the interaction of obliquely incident planewaves with doubly periodic structures.  We employed a field transformation to provide a formulation free from the well-known causality issues with periodic boundary conditions in time.  The field transformations were applied to the first order Maxwell's Equations, and a numerical flux was derived using an equivalent set of transformed equations.  The computational framework was validated using existing results in the literature.  While the particular examples elaborated in this paper employed a planewave ABC, we are currently developing an exact time domain Floquet radiation boundary condition.  Future applications include the optimization of photonic band gap structures and complex frequency selective surfaces.

\section{Acknowledgment}

This work was supported by the National Science Foundation through grant CCF:1018576.  The authors would like to thank General Electric (GE) for support, and acknowledge computing support from the HPC Center at Michigan State University, East Lansing.

\bibliographystyle{unsrt}
\bibliography{./Periodic_DG_Paper}{}

\end{document}